\documentclass[11pt,titlepage]{article}
\usepackage{setspace} 
\usepackage[round,numbers,sort&compress]{natbib} 
\usepackage{graphics}
\usepackage[english]{babel}
\usepackage{graphicx}
\usepackage{amsmath}
\usepackage{amssymb}
\usepackage{wrapfig}
\usepackage{subfig}
\usepackage{float}
\usepackage{color}

\newcommand{\Ca}{$Ca^{2+}$}
\newcommand{\sol}{\rho_{sol}}
\newcommand{\gel}{\rho_{gel}}
\newcommand{\rv}[1]{\boldsymbol{#1}}
\newcommand{\vn}{\rv{\nabla}}

\title{An active poroelastic model for mechanochemical patterns in protoplasmic droplets of Physarum polycephalum}

\author{
	Markus Radszuweit\thanks{ 
           Corresponding author. Address:
	   Hardenbergstrasse 36,
	   10623 Berlin, Germany
	   Tel.:~(+49-30) 314-27681} \\
	   Mathematische Modellierung und Datenanalyse, \\
	   Physikalisch Technische Bundesanstalt, Berlin, Germany \\ 	
	\and Harald Engel \\
	Institut f\"ur theoretische Physik, \\
	Technische Universit\"at Berlin, Berlin, Germany  \\	  
	\and Markus B\"ar \\
	    Mathematische Modellierung und Datenanalyse, \\
	   Physikalisch Technische Bundesanstalt, Berlin, Germany
}

\date{\today}

\begin{document}

\maketitle

\abstract{
Motivated by recent experimental studies, we derive and analyze a two-dimensional model for the contraction patterns observed in protoplasmic droplets of Physarum polycephalum. 
The model  couples a model of an active poroelastic two-phase medium with equations describing the spatiotemporal dynamics of the intracellular free calcium concentration.
The poroelastic medium is assumed to consist of an active viscoelastic solid representing  the cytoskeleton and  a viscous fluid describing the cytosol. 
The model equations for the poroelastic medium are obtained from 
continuum force-balance equations that include the relevant mechanical fields and an incompressibility relation for the two-phase medium. 

The reaction-diffusion equations for the calcium dynamics in the protoplasm of Physarum are extended by advective transport due to the flow of the cytosol generated by mechanical stresses. 
Moreover, we assume that the active tension in the solid cytoskeleton is regulated by the calcium concentration in the fluid phase at the same location, which introduces  a chemomechanical feedback.
A linear stability analysis of the homogeneous state without deformation and cytosolic flows exhibits an oscillatory Turing instability for a large enough mechanochemical coupling strength.
Numerical simulations of the model equations reproduce a large variety of wave patterns, including traveling and standing waves, turbulent patterns, rotating spirals and antiphase oscillations in line with experimental observations of contraction patterns in the protoplasmic droplets.

}

\clearpage


\section*{Introduction}

The true slime mold \textit{Physarum polycephalum} is an extensively studied system in biophysics. 
The plasmodial stage is of particular interest, since it exhibits, despite the relatively 
simple organization of this unicellular organism, seemingly ``intelligent'' physiological processes \citep{UED05}. 
In this context the term ``intelligent'' means that, given an external stimulus, the plasmodium 
optimizes its cell shape, vein network and growth with respect to transport efficiency, robustness with respect to link deletion and avoidance of unfavorable conditions. 
Recent experiments along these lines show that plasmodia were able to reproduce public transport networks on the scale of a petri dish \citep{TER10} and to ``solve'' maze problems such as finding the shortest path 
between two food sources placed at the exits of a labyrinth \citep{NAK00}. 
Several groups have also investigated the topology and dynamical evolution of the vein network in large Physarum plasmodia with graph theoretical and statistical physics tools 
\citep{BUH10,FOB12,BH13}.
A second remarkable phenomenon is the synchronization of the contraction patterns in the tubular vein network that generates shuttle streaming to distribute nutrients efficiently throughout the organism \citep{KAM81}. 
From the perspective of biophysics it is natural to consider these phenomena in the framework of self-organized complex systems \citep{NAK08}. 
For the formulation of mathematical models a basic 
understanding of chemical and mechanical processes in the protoplasma is needed.\\

A first model for strand contraction combined the viscoelastic properties of the ectoplasmic wall with a reaction kinetics that regulates the contractile tension of the actomyosin 
system \citep{OST84,TEP91}. 
Later, several models in the form of reaction-diffusion (RD) \citep{TER05} and reaction-diffusion-advection equations (RDA) \citep{NAK98,YAM07} were formulated that use homogenized quantities, for 
instance the average strand thickness. 
These models describe Physarum protoplasma as an oscillatory medium and treat the mechanical feedback in an oversimplified, qualitative  way.
More realistic models consider, instead, a two-phase description that distinguishes a fluid sol (= cytosol) and a solid gel (= cytoskeleton) phase. 
Some of these models  account for sol-gel transformations and were used to explain flow-channel formation \citep{GUY11} and front dynamics \citep{UEK11}.\\ 

\begin{figure}[htbp]
   \begin{center}
      \includegraphics*[width=0.8\textwidth]{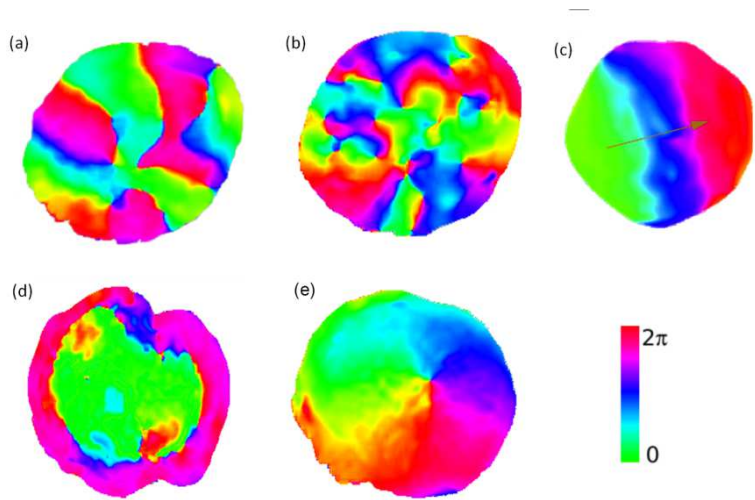}
      \caption{Contractions patterns: Experiments with protoplasmic droplets of Physarum polycephalum \citep{TAK08,TAK10}. The color represents the local phase of oscillation obtained by a Fourier transformation of the spatiotemporal 
	height data: a) standing wave, b) many irregular spirals, c) traveling wave, d) antiphase patterns, and e) single spiral.}
      \label{fig:exp}
   \end{center}
\end{figure}

Experiments with microplasmodia, i.e. small plasmodia of sizes ranging from $100\mu m$ to several millimeters, provide a possibility to study internal amoeboid dynamics  of Physarum without the pronounced vein structures usually present in Physarum cells of larger size. 
Such microplasmodia 
are produced by extracting cytosol from a Physarum vein and placing it on a substrate. 
Given a sufficient amount of cytosol, protoplasmic droplets will reorganize and form a new independent cellular entity. 
During the first hours of this process such  cells show a surprising wealth of spatiotemporal mechanical contraction patterns \citep{TAK08,TAK10,SH10}.
 The fact that the cell morphology does not change dramatically and that the cell do not migrate during the 
first hours, permits observation of the mechanical deformation patterns and waves in a quasi-stationary setting . 
The observed patterns include spirals, traveling and standing waves as well as antiphase oscillations (see Fig. \ref{fig:exp}).\\
Various patterns were reproduced previously by a qualitative particle-based model \citep{TSU11}. However, this descripted provided no information about the mechanical quantities that are essential to understandthe intracellular deformation waves and patterns seen in the experimentes. \\

In a more general context, studying the spatiotemporal instabilities 
and the related symmetry breaking in intracellular processes has become an important
tool to understand many biological processes. 
In a pioneering paper \citep{TUR52}, Turing suggested that the interplay of 
reactions and diffusion processes provides a fundamental mechanism for morphogenesis.  
More recently, 
the reaction-diffusion mechanism proposed by Turing was 
applied to pattern formation in single cells and shown to 
be relevant, e.g., for cell polarity \citep{STRI07} and the control of 
cell division \citep{JIL11}. 
In addition, models for intracellular pattern formation that include mechanical forces and the resulting 
advection processes have moved into the focus of biophysical research \citep{BOI11,HOW11}. 
Transport in cells such as the Physarum droplets described above does not only occur by passive diffusion 
\citep{NAK08}, it is often actively driven by stress generation 
from cytoskeletal filaments \citep{BET11,OST84a}. 
The cytoskeleton is an active cellular material in the form of a network of filaments \citep{MCK10}. 
Active molecular motors control the mechanical 
properties of this network and keep the system 
far from thermodynamic equilibrium.
Active gel models describe the cytoskeleton as an active viscous fluid \citep{JOA07}.
In contrast, experiments on inhomogeneous hydration 
in cells, where large pressure gradients in the cell are observed 
\citep{CHA09} indicate that the cytoplasm can behave like a 
porous elastic sponge-like solid (cytoskeleton) penetrated 
by a viscous fluid phase (cytosol) \citep{MIT08,MOU13}.


%
Moreover, several multiphase flow models have been proposed as appropriate description of cytoplasmic dynamics \citep{DEM86,COG10}. 

In this paper, we derive and investigate a 
poroelastic two-phase model of the cytoplasm assuming 
a viscoelastic solid phase and a fluid phase.
Furthermore, we incorporate an active tension in the solid phase which 
is regulated by the concentrations of free calcium ion in the fluid phase (cytosol), that 
are in turn advected by the fluid phase. 
To account for the calcium oscillation observed in experiments with Physarum, 
a simple active poroelastic model derived earlier \citep{RAD13} is extended by a coupling to an oscillatory reaction-diffusion dynamics of the intracellular calcium concentration \cite{SMI92}.
The choice of the poroelastic approach is motivated by the fact that the typical
oscillation period of 1 - 2 minutes connected with the spatiotemporal patterns discussed above 
is considerable shorter than the experimentally observed time of 
3 - 6 minutes at which the cytoskeleton starts to exhibit fluid behavior \cite{NAG78}. 
Hence, the resulting model describes the cytoskeleton as an active 
viscoelastic solid coupled to a passive fluid in contrast 
to earlier works that had modelled the cytoskeleton itself as an 
active fluid \citep{BOI11} addressing long time scales, for which fluidization of the cytoskeleton 
has already occured.  \\

The  inclusion of the calcium oscillator is necessary because it is known to be essential in the regulation of the contractile actomyosin system \citep{YOS81}.
Altogether, in article  we derive  and analyze a model for the intracellular dynamics of protoplasmic droplets that treats the cellular mechanics in the framework of a continuous two-phase active poroelastic modelcoupled to an oscillatory biochemical medium.  
As a consequence of internal 
pressure gradients a flow of cytosol occurs that will be included as a feedback to the chemical part of the system. 
Together with the calcium oscillator proposed in \citep{SMI92} we introduce an reaction-diffusion-advection (RDA) equation for the concentration of calcium.\\ 
%

%
%
%

In the methods section we introduce and derive the mathematical  mode with a description divided into a mechanical and a biochemical part. 
Subsequently,the choice of physical parameters introduced in the model is discussed and   the numerical methods used to discretize and solve the PDEs is described. 
The next section contains the results obtained by linear stability analysis at the homogeneous steady state (HSS) and a two-parameter phase diagram with 
numerical simulations. 
We present also a selection of qualitatively different patterns obtained from simulations of our model and compare them to earlier experimentally findings and demonstrate that the variety of patterns found in the experiments with Physarum droplets are reproduced successfully. 
The paper is conclueded with a discussion of the presented model, its limitations and possible extensions.


\section*{Materials and Methods}

\subsection*{Model: Mechanical part}

\begin{figure}[htbp]
   \begin{center}
      \includegraphics*[width=0.8\textwidth]{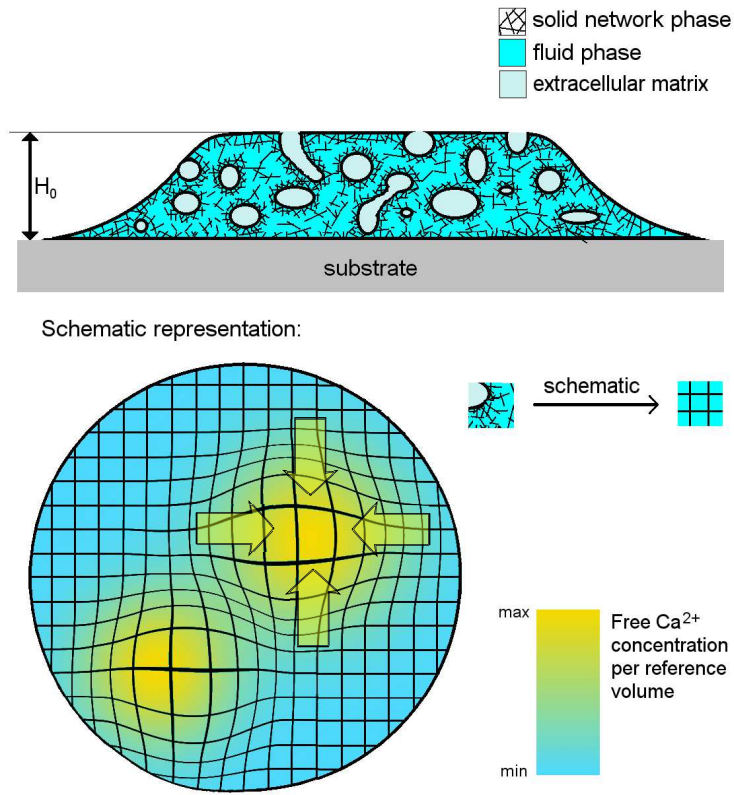}
      \caption{Schematic representation of the the two-phase model: Drawing of a Physarum microdroplet (top) in side view showing the plasmalemma invaginations (light blue), the fluid phase of the cytoplasm (blue) and the solid network 
       phase (black). Top view of the droplet in the simplified framework of our two-phase model (bottom). Deformations of the cytoskeleton are represented by a distorted grid, flow field in the cytosol by arrows and free \Ca concentration in yellow.}
      \label{fig:droplet}
   \end{center}
\end{figure}

Physarum protoplasm contains a rudimentary form of an actomyosin system that is also present in cells of higher vertebrates.
In contrast to muscle cells the actin filaments in Physarum are randomly oriented in the cortex \citep{NAG75,BRX87}.
In our model, we assume that the cytoplasm contains a solid filamentous phase (gel phase) that has viscoelastic properties and exerts contractile tension on the system. 
The derivation given below is analogous to a recently published generic model for active poroelastic media \citep{RAD13}. 
The fluid part of the cytoplasm is modeled as a passive fluid (sol phase) that permeates the cytoskeleton \citep{CHA09}. 
The velocity field in the sol phase will be expressed by the variable $\rv{v}$. 
Typical Reynolds numbers that arise from the cytoplasmic flow are small ($Re\ll 1$) and inertia is negligible.
The volume fraction of solid material is denominated as $\gel$ and the fraction of the fluid material as $\sol$ with the additional constraint $\sol+\gel=1$. \\

We define a body reference coordinate system $\rv{x}$ and a displacement field $\rv{u}$ that gives the deviation of the deformed coordinates $\rv{X}$: 
$\rv{u}(\rv{x},t)=\rv{X}(\rv{x},t)-\rv{x}$ at a time $t$. 
We assume only small deformations and thus restrict ourselves to linear elastic theory. 
The gel velocity is the substantial time derivative $\dot{\rv{u}}=\partial_t\rv{u}+(\vn\rv{u})\cdot\dot{\rv{x}}$ of the displacement field. 
Since the gel is fixed in the reference coordinate system ($\dot{\rv{x}}=0$), the substantial time derivative $\dot{\rv{u}}$ is identical to the partial time derivative $\partial_t\rv{u}$.
\\

To determine the flow and displacement field we consider force-balance equations of the form
\begin{eqnarray}
 \vn\cdot(\gel(\rv{\sigma}_{\textrm{gel}}^{\textrm{pass.}}+T_a\mathbf{1}))+\rv{f}_{\textrm{gel}} & = 0
 \label{eq:fb_gel} \\
 \vn\cdot(\sol\rv{\sigma}_{\textrm{sol}}^{\textrm{pass.}})+\rv{f}_{\textrm{sol}} & = 0,
 \label{eq:fb_sol}
\end{eqnarray}
where the passive sol and gel stresses $\rv{\sigma}_{\textrm{sol/gel}}^{\textrm{pass.}}$ are determined by linear constitutive laws and $\rv{f}_{\textrm{sol/gel}}$ are force densities. 
We assume that the gel phase 
is a porous viscoelastic active material \citep{RAD13} that is able to exert contractile stresses by interaction of the myosin-motor system with the actin filaments. 
The filament orientation in the cortex of Physarum that mainly 
determines the active and passive properties of the medium is random \citep{BRX87}. 
Hence, we use an isotropic constitutive law for the elastic stress-strain relation.
Isotropy is also supposed for the active stress $\rv{\sigma}_{\textrm{gel}}^{\textrm{act.}}=T_a\mathbf{1}$ that is added 
to the passive stress in Eq. (\ref{eq:fb_gel}). 

We neglect the influence of the solation and gelation dynamics of the filamentous on the viscoelastic parameters introduced above.\\

It has been suggested to consider the cytoplasm as an incompressible medium \citep{ALT99}. 
In the three-dimensional bulk the total mass flux must be zero and for small strains can be expressed as \citep{DEM86}:
\begin{equation}
 \vn\cdot(\gel\dot{\rv{u}}+\sol\rv{v})=0. 
 \label{eq:inc}
\end{equation}
In the reduced two-dimensional model here this condition is generally violated. 
Since we assume small deformations only, this effect is, however, small and will be neglected in the following. 
In the following we assume constant sol and gel fractions $\sol^0$ and $\gel^0$ throughout the medium.
This is justified, because we consider only small deformations. 
As a result the transport of cytosol and the related potential inhomogeneities of the fields $\sol$ and $\gel$ lead only to second-order corrections in the mechanical equations (for details see supplementary material). \\

We include a hydrostatic pressure $p$ into the stress tensors of sol and gel that originates from the incompressibility of the material expressed by Eq. (\ref{eq:inc}) and neglect the osmotic pressures caused 
by a difference in the chemical potential (see supplementary material). 
We assume a Kelvin-Voigt viscoelastic constitutive law (see e.g. \citep{BAS11}) for the gel phase. 
Using Darcy's law $\rv{w}=-\frac{k}{\eta}\vn p$ the following relation for the drag force is obtained

\begin{equation}
\rv{f}_{\textrm{gel}}=-\rv{f}_{\textrm{sol}}=\sol^0\gel^0\beta\rv{w},
\end{equation}

where the parameter $\beta$ is the ratio between the dynamic viscosity $\eta$ of the cytosol and the permeability $k$ of the porous medium.
Instead of the fluid velocity $\rv{v}$ in the laboratory frame, we need to consider the velocity $\rv{w} =\rv{v}-\dot{\rv{u}}$ in the body-reference frame introduced above. 
The sol phase is considered as a passive Newtonian fluid. 
Hence, only viscous stresses are included for the sol phase.
As a result, Eq. (\ref{eq:fb_sol}) corresponds to the Brinkman equation \citep{BRI49}. 
With the usual expression $\rv{w}=-\frac{k}{\eta}\vn p$ for Darcy's law and Eq. (\ref{eq:fb_sol}) one can relate the 
coefficient $\beta$ in our model to the viscosity $\eta$ of the cytosol and the permeability $k$ of the medium: $\beta\gel^0=\eta/k$.
The mechanical force balance equations in the final form (for a detailed derivation we refer the reader to the supplementary material)
together with the incompressibility condition read:
\begin{equation}
 \begin{array}{ll}
 \vn\cdot(\rv{\sigma}_{gel}^{dis}+\rv{\sigma}_{gel}^{el}+(T_a-p)\mathbf{1})+\beta\sol^0(\rv{v}-\dot{\rv{u}}) & =0 \\
 \vn\cdot(\rv{\sigma}_{sol}^{dis}-p\mathbf{1})-\beta\gel^0(\rv{v}-\dot{\rv{u}}) & =0 \\
 \vn\cdot(\gel^0\dot{\rv{u}}+\sol^0\rv{v}) & =0.
 \end{array}
 \label{eq:mech_full}
\end{equation}

\subsection*{Model: Chemical part}

Calcium ions play a key role in the regulation of contraction in Physarum polycephalum.
To describe the calcium kinetics, we use a the biochemically realistic model presented by Smith and Saldana \citep{SMI92}.
This model describes an oscillation mechanism that is driven by a phosphorylation-dephosphorylation cycle of myosin light chain kinases.
A crucial feature of the Smith-Saldana model are autonomous calcium oscillations that occur already in the absence of mechanical feedback. 
Another entirely different mechanism was introduced to explain oscillations of plasmodial strands of Physarum, where the necessary feedback for the oscillations is provided by mechano-sensitive channels.
This feedback acts upon a non-oscillatory calcium reaction kinetics and is therefore required to obtain the oscillations \citep{TEP91,ROM95}. 
This model, however, predicts that the free calcium concentration and the mechanical tension oscillate in phase, whereas experiments exhibit an antiphase oscillation with a phase shift of $\pi$ between these two quantities \citep{YOS81}.
Moreover, experiments in the homogenate of Physarum plasmodium wherein deformations and mechanical stresses are not possible yield calcium oscillations giving further evidence for an autonomous calcium oscillator
\citep{YOK82}. 
The Smith-Saldana model \citep{SMI92}, in contrast, provides the correct phase shift between calcium and tension. 
It can be reduced to two ordinary differential equations involving the free calcium concentration $n_c$ and the fraction of phosphorylated myosin-light-chain kinases $\phi$:
\begin{equation}
\begin{array}{lll}
 \dot{n}_c & = & \left(k_L(N_c-n_b-n_c)-k_Vn_c-\frac{\partial n_b}{\partial\phi}(k_Q(n_c) (1-\phi)+k_E\phi)\right)/\left( 1+\frac{\partial n_b}{\partial n_c}\right )\\
  & =: & f_c(n_c,\phi) \\
 \dot{\phi} & = & k_Q(n_c) (1-\phi)+k_E\phi=:f_\phi(n_c,\phi).
 \label{eq:chem}
\end{array} 
\end{equation}
These equations involve the functions:
\begin{equation}
 \begin{array}{ll}
  k_Q(n_c) = & k_Q^0\left (\frac{K^\star n_c}{1+K^\star n_c}\right )^\beta\\
  n_b(n_c,\phi) = & 2N_M\left (\frac{K_an_c}{1+K_an_c}(1-\phi)+\frac{K_bn_c}{1+K_bn_c}\phi)\right ).
 \end{array}
\end{equation}
The dependence of the dephosphorylation rate $k_Q$ on $n_c$ originates from a signal cascade (see \citep{SMI92} for details) and the second equation for $n_b$ expresses the myosin-bound calcium. 
For convenience, the right-hand sides of the ODEs 
are abbreviated as follows: $\dot{n}_c=f_c(n_,\phi)$ and $\dot{\phi}=f_\phi(n_c,\phi)$. 
This model allows to relate the state of the calcium oscillator directly to the fraction $\theta$ of myosin molecules 
that are activated to bind with actin:
\begin{equation}
 \begin{array}{ll}
   \theta(n_c,\phi) = & \frac{k_P(1-q_{2a}(n_c))}{k_D+k_P(1-q
_{2a}(n_c))}(1-\phi)+\frac{k_P(1-q_{2b}(n_c))}{k_D+k_P(1-q_{2b}(n_c))}\phi \\
    q_{2a/b}(n_c) = & (K_{a/b}n_c)^2/(1+K_{a/b}n_c)^2.    
 \end{array}
 \label{eq:theta}
\end{equation}
Then, the model is completed by a third equation that relates the activated fraction of myosin to the active tension $T_a$. 
In contrast to the original work of Saldana and Smith, we introduce a relaxation equation analogous to models of cardiac myocytes \citep{PAN07}:
\begin{equation}
 \dot{T}_a=(F_T\theta(n_c,\phi)-T_a)/\tau_T,
 \label{eq:tension}
\end{equation}
where $\tau_T$ is the relaxation time the tension needs to approach its equilibrium value and $F_T$ represents the mechanical coupling strength.\\

The bifurcations of the ODE system Eqs. (\ref{eq:chem}) and (\ref{eq:tension}), were analyzed in \citep{RAD10}. 
It was shown that a supercritical Hopf bifurcation occurs. \\

The next step in the derivation of the model 
is a spatial extension of Eq. (\ref{eq:chem}) to a reaction-diffusion-advection (RDA) system. 
The only species in this model that is transported by diffusion is the free calcium $n_c$. 
Altogether,  the following equation is obtained
\begin{equation}
  \partial_t n_c+\vn\cdot(n_c\rv{w}) = D_c\Delta n_c+f_c(n_c,\phi), 
 \label{eq:adv}
\end{equation}
where  the fluid velocity in the body reference frame introduced above $\rv{w}$ and the diffusion constant of the free calcium in the cytosol is $D_c$. Note that Eq. (\ref{eq:adv}) is given 
in the body-reference coordinate sytem.

\subsection*{Summary of the model}
Collecting all pieces of the model described so far, we obtain the following PDE system: 
\begin{eqnarray}
 0 & = & \eta_{gel}^{shear}\Delta\dot{\rv{u}} +\eta_{gel}^{bulk} \vn(\vn\cdot\dot{\rv{u}})+G\Delta\rv{u} \label{eq:gel}\\
   &   & +K\vn(\vn\cdot\rv{u})+\vn(T_a-p)+\sol^0\beta(\rv{v}-\dot{\rv{u}}) \nonumber \\
 0 & = & \eta_{sol}^{shear}\Delta\rv{v}+\eta_{sol}^{bulk} \vn(\vn\cdot\rv{v})-\vn p-\gel^0\beta(\rv{v}-\dot{\rv{u}}) \\
 0 & = & \vn\cdot(\gel^0\dot{\rv{u}}+\sol^0\rv{v}) \label{eq:inc0} \\
 \partial_t {T}_a &=& (F_T\theta(n_c,\phi)-T_a)/\tau_T \label{eq:Ta} \\
 \partial_t {n}_c &=& -\vn\cdot(n_c(\rv{v}-\dot{\rv{u}}))+D_c\Delta n_c+f_c(n_c,\phi) \label{eq:nc}\\
 \partial_t {\phi} &=& f_\phi(n_c,\phi).\label{eq:phi}
\end{eqnarray}
Here, we have introduced the shear and bulk viscosities $\eta_{sol/gel}^{shear}$ and $\eta_{sol/gel}^{bulk}$ of sol and gel phase and the linear elastic shear and compression 
modulus of the gel phase $K$ and $G$. 
Note that the sol fraction and the free calcium concentration are given in the body-reference frame.\\
The above equation are defined in a circular domain that mimicks the 
geometry of the Physarum droplets in experiments \citep{TAK08,TAK10,SH10}. 
Dirichlet conditions are imposed at the boundary of the domain both for the displacement field, i. e. $\rv{u}(\rv{x},t)=0$, and for the flow field , i. e. $\rv{v}(\rv{x},t)=0$.  
Assuming, a non-permeable membrane of the droplet, no-flux boundary conditions $\vn n_c(\rv{x},t)\cdot\rv{n}(\rv{x},t)=0$ are employed for the calcium concentration $n_c$,  where $\rv{n}$ is the normal vector at the boundary.
For these boundary conditions the pressure field is determined up to an arbitrary constant, that is fixed by setting $p(\rv{x})=0$ at a given position in the system.
As initial conditions we apply a small random noise $\xi$ 
with zero mean as a perturbation to the homogeneous steady state solution.\\

To compare with the experimentally measured height profiles of the droplet, one needs to estimate the local height field $H(\rv{x},t)$ that does not appear explicitly in the two-dimensional model. 
The relative height deviation is defined as $h(\rv{x},t):=(H(\rv{x},t)-H_0)/H_0$.
We relate this relative height deviation to the divergence of the displacement field computed in the two-dimensional model:
\begin{equation}
  h(\rv{x},t)\propto \vn\cdot\rv{u}(\rv{x},t).
\end{equation}
This approximation is based on the idea that deformations are locally isotropic \cite{BRX87}.

\subsection*{Parameters}

A summary of the parameters used in the model is given in Table \ref{tab:params}. 
The values for the chemical oscillator are taken from Ref. \citep{SMI92}. 
For the affinity $K_a$ two different values are considered.
Autonomous calcium oscillations are obtained for $K_a = 2.3 \mu M^{-1}$,
whereas a stationary calcium concentration is found for $K_a = 2.0 \mu M^{-1}$ in the absence of mechanical feedback. 
For effect of the parameter $K_a$ on the autonomous calcium oscillator model for Physarum was studied systematically mechanical in Refs. \citep{SMI92,RAD10}. 
For the diffusion coefficient of the free calcium we use a typical value of $D_c=0.03$ $mm^2/min$ for small ions in cytoplasm \citep{DON87}. 
In Physarum, the percentage of actin in the cytoplasm is estimated to be in the range $15-25\%$ \citep{KES76}. 
Since other proteins also contribute to the solid gel phase, we set $\gel^0=0.25$ and $\sol^0 = 0.75$. 
The Young modulus $E$ of a Physarum strand was determined to be of the order $\approx 10$ $kPa$ \citep{NOR40}. 
For sponge-like materials, measurements show that the Poisson ratio is very low: $\nu\approx 0$ \citep{BOR04} implying $G \approx K$. 
Therefore, we have set the parameters $G=K=8.9$ $kPa$.
 A typical value for the generated tension in plasmodial strands is 
$T_a\approx 20$ $kPa$ \citep{WOH77}. 
According to Eq. (\ref{eq:Ta}), the active tension $T_a$ relaxes to an
equilibrium value of $F_t \theta$.
Because $\theta_{max}<0.1$ (see \citep{rad-phd-2013}, values up to $F_T=350$ $kPa$ for $F_T$ have been considered. 
Moreover, we have varied $F_T$ in our study to illustrate the influence of the strength of mechanical coupling on the pattern dynamics. 
For the dynamic sol viscosity $\eta^{shear}_{sol}$ in Physarum, values in the range of $0.1-0.5$ $Pas$ where measured \citep{SAT83}. 
Alternatively, one can calculate the 
sol viscosity from velocity profiles measured in Physarum \citep{BYK09}.
With this method we obtain a dynamic sol viscosity of around $10$ $Pas$.
In the model, the value of the sol viscosity is then set to 
$\eta^{shear}_{sol}=1$ $Pas$.
There are no measurements for the viscous shear damping $\omega\eta^{shear}_{gel}$ for the filament phase in Physarum. 
We refer instead to investigations of a similar system of a composite network containing actin filaments and microtubuli \citep{PEL09}. 
Therein, the frequency dependent complex dynamic shear modulus $G(\omega)=G'(\omega)+iG''(\omega)$ was measured. 
For a viscoelastic material 
described by the Kelvin-Voigt model, one identifies $G''(\omega)=\omega\eta^{shear}_{gel}$. 
For a typical frequency $\omega=2\pi/min^{-1}$ of calcium oscillations in Physarum a value of $\eta^{shear}_{gel}\approx 1$ $Pas$ is obtained.
The sol and gel phases itself are incompressible and thus, no bulk parameters can be measured. 
Here, we neglect the influence from the bulk viscosities and set $\eta_{sol/gel}^{bulk}=0$. 
Since the permeability $k\propto \ell_{pore}^2$, the drag coefficient is $\beta\propto\eta_{sol}/\ell_{pore}^2$, where $\ell_{pore}$ is the average pore size. 
Porous structures in Physarum exhibit several spatial scales: The actin-network pore size $\ell_{pore}=0.25\mu m$ 
\citep{NAG75} and the membrane structure of invaginations posesses pores with a typical diameter of ca. $10\mu m$ \citep{BRX87}. 

If one assumes that larger porous structures determine the permeability,  a drag coefficient of $\beta\approx 10^4$ $kg/(mm^3min)$ is obtained. 
Since, however, the porous structure of the Physarum cytoskeleton may vary over time or depending on the developmental state, we have varied the parameter $\beta$ over a range from $\approx 6\cdot 10^2-9.6\cdot10^5$ $\frac{kg}{mm^3min}$.

\subsection*{Linear Stability Analysis}
Linear stability analysis is used here to investigate the spatiotemporal instability near the homogeneous steady state (HSS). 
The HSS of the system is given by $\rv{u}^\star\equiv\rv{v}^\star\equiv 0$, $p^\star=const.$ and the steady state values of chemical components and active tension are given by the implicit equations $f_c(n_c^\star,\phi^\star)=0$, $f_\phi(n_c^\star,\phi^\star)=0$ and $T_a^\star=F_T\theta(n_c^\star,\phi^\star)$.
Note, that this HSS corresponds to state without deformation and without any fluid motion.
The growth rate and the frequency (for imaginary eigenvalues) of a small perturbation $(\Delta n_c,\Delta\phi,\Delta T_a,\Delta\rv{u},\Delta\rv{v},\Delta p)e^{i\rv{q}\rv{x}+\lambda t}$ to the HSS is given by the dispersion relation $\lambda(\rv{q})$.

\begin{figure}[htbp]
\begin{minipage}{\textwidth}
\begin{center}
 a) \includegraphics*[width=2.835in]{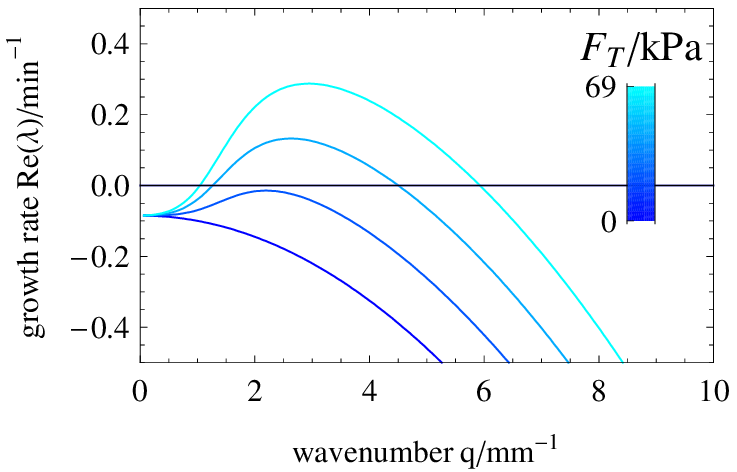}
 \includegraphics*[width=1.8in]{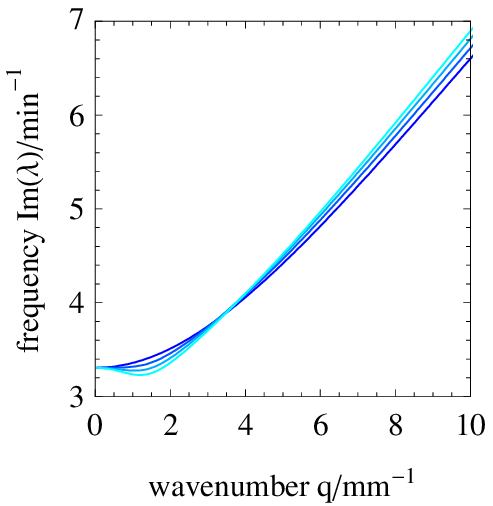}
\end{center}   
\end{minipage}

\begin{minipage}{\textwidth}
\begin{center}
 b) \includegraphics*[width=2.835in]{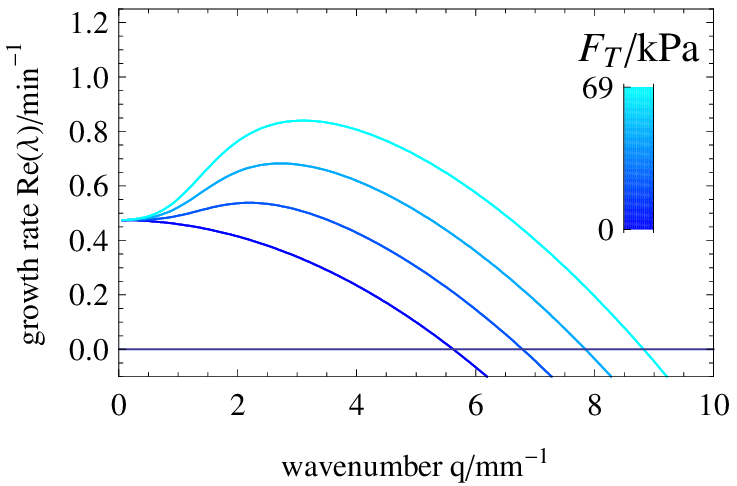}
 \includegraphics*[width=1.8in]{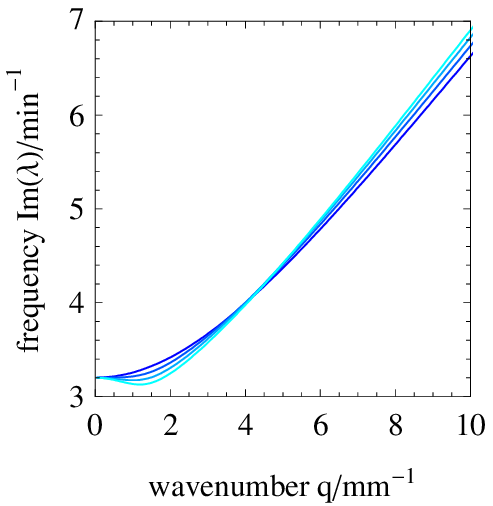}
\end{center}
\end{minipage}
\caption{Real part (left) and imaginary part (right) of the branch of the dispersion relation with largest real parts of the eigenvalues for a) stable HSS ($K_a=2.0$ $\mu M^{-1}$) and b) unstable HSS ($K_a=2.3$ $\mu M^{-1}$). The imaginary part is nonzero for all unstable wavenumbers: $Im(\lambda(q))\neq 0$. The mechanical coupling strength $F_T$ increases from dark blue to cyan. The drag coefficient is chosen to be $\beta=5\cdot 10^4$ $kg/(mm^3min)$ and the remaining paramters are given in Table \ref{tab:params}.}
\label{fig:disp_FT}
\end{figure}

\subsection*{Numerical Integration}
Numerical simulations of the full nonlinear dynamics of the above model were performed to obtain solutions of the system of Eqs. (\ref{eq:gel})-(\ref{eq:phi}). 
The integration was done with a hybrid method consisting 
of a finite element (FEM) and finite volume discretization scheme(FVM). 
Two-dimensional meshes were generated with the free software \textit{Triangle} \citep{SHE96}. 
With the operator-splitting technique the time integration step was divided into substeps that are carried out with different types of solvers: a fourth order Runge-Kutta method for the nonlinear reaction part, a linear FEM with implicit Euler stepping for the parabolic and elliptic parts \citep{note1} and a FVM 
using the dual mesh of the triangulation (Voronoi diagram) for the advection step. 
The number of nodes for the mesh discretizing a disc with radius $r=1$ $mm$ was $N=5218$. The time step size $\Delta t$ was chosen to be 
$=0.01 min$ and the typical simulation length was $100$ $min$.


\section*{Results}

\subsection*{Linear stability analysis and dispersion relation}

\begin{figure}[htbp]
   \begin{center}
    a)
     \includegraphics*[width=0.7\textwidth]{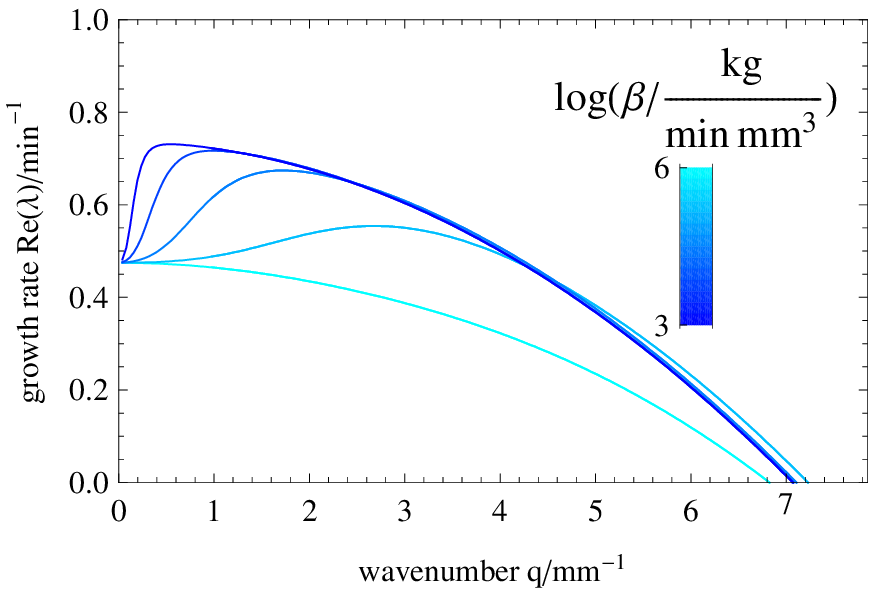}\\
    b)
     \includegraphics*[width=0.7\textwidth]{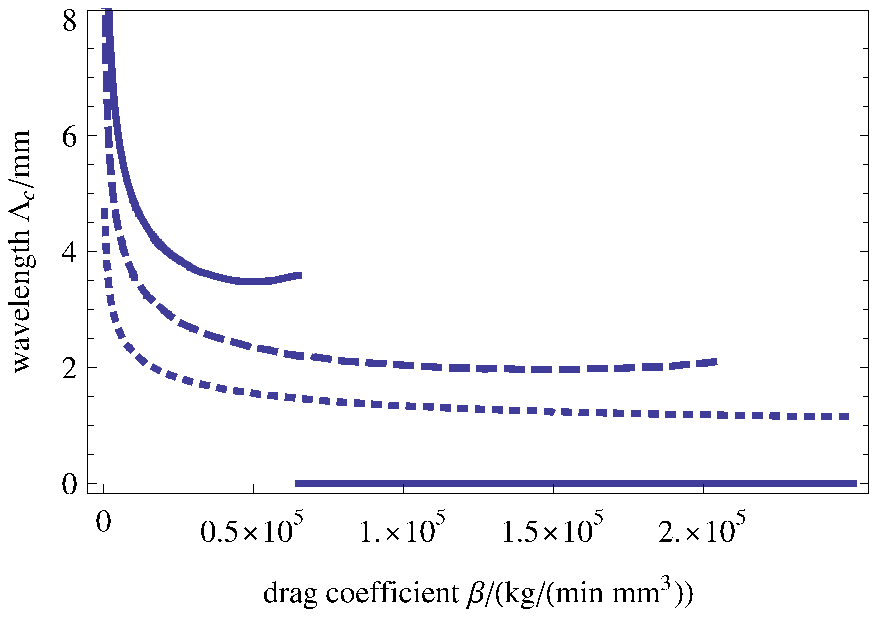}    
      \caption{Branch of the dispersion relation with largest real part a) for different drag coefficients $\beta$ that range over three orders of magnitude (logarithmic scale, from dark blue to cyan). The plots show only the real part. 
      The mechanical coupling strength is kept constant at $F_T=28$ $kPa$. b) Wavelength $\Lambda_c=2\pi/q_c$ of the fastest growing mode versus drag coefficient $\beta$ for three different mechanical coupling strength $F_T=14$ 
      $kPa$ (solid line), $F_T=44$ $kPa$ (dashed line) and $F_T=140$ $kPa$ (dotted line). 
      The remaining paramters are given in Table \ref{tab:params}.}
      \label{fig:disp_k}
   \end{center}
\end{figure}

\begin{figure}[htbp]
   \begin{center}
    \begin{minipage}{2.94in}
     \includegraphics*[width=\linewidth]{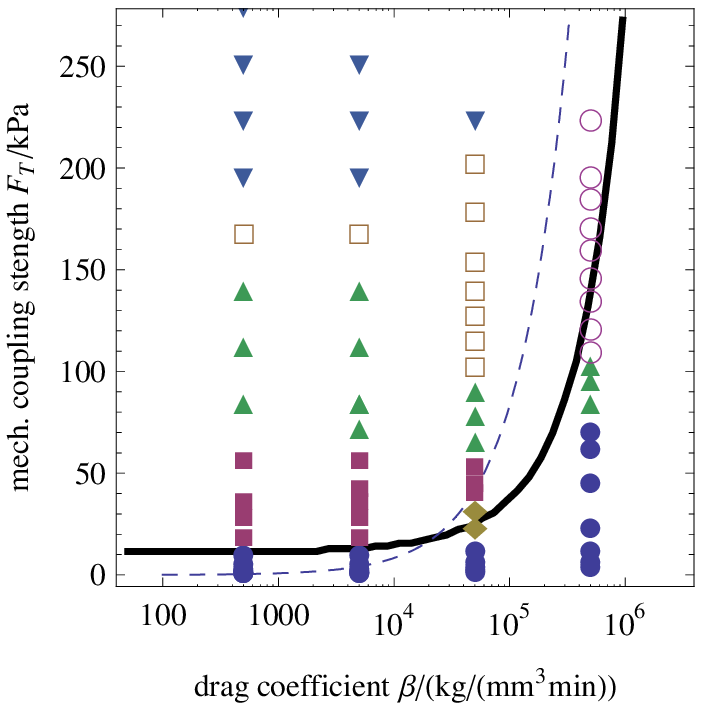}
    \end{minipage}
    \begin{minipage}{1.125in}
     \includegraphics*[width=\linewidth]{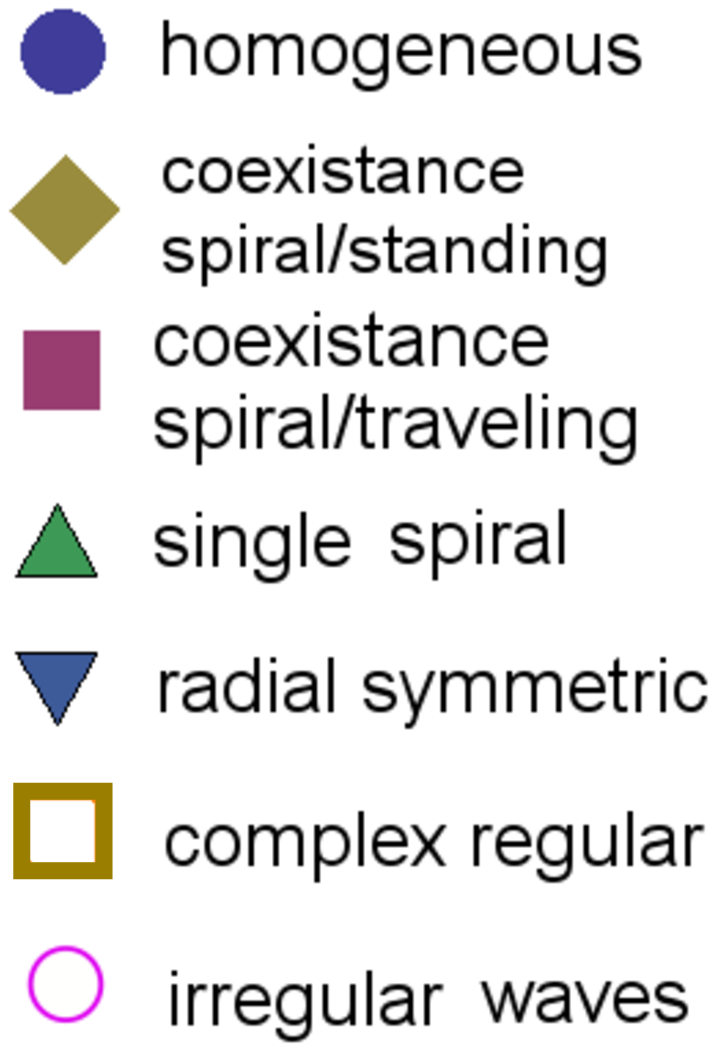}
    \end{minipage}
   \end{center}
   \caption{Phase diagram in the plane spanned by the mechanical coupling strength $F_T$ and the drag coefficient $\beta$. The black line denotes the threshold coupling strength $F_{T}^{thr}$, where the most unstable 
    mode according to linear stability analysis of the HSS is nonzero. The dashed blue curve separates the plane according to Eq. \ref{eq:Pe} in regions with $Pe>1$ (larger $F_T$) and $Pe<1$ (smaller $F_T$).}
   \label{fig:phase_diag}
\end{figure}
Here ,the mechanical coupling strength $F_T$ and the drag coefficient $\beta$ are varied and to reveal their influence on the stability of the steady and the shape of the dispersion curve $\lambda(\rv{q})$. 
In Fig. \ref{fig:disp_FT} the branch of eigenvalues
with the largest real part is shown for different $F_T$. 
The real and imaginary parts of the dispersion relation are displayed in separate plots. 
Fig. \ref{fig:disp_FT} a) shows the case where the HSS is stable in absence of mechanical coupling ($F_T = 0$). 
If $F_T$ is increased above a critical value, the HSS exhibits a wave instability (oscillatory Turing instability).
In Fig. \ref{fig:disp_FT} b) we consider a case where 
the HSS is already unstable against oscillations for $F_T = 0$, i.e. $Re(\lambda(0))>0$. 
For larger wavenumbers $q$ of the perturbation the imaginary part of the growth rate increases indicating that the frequency of waves is larger than the frequency of homogeneous oscillations.\\

In the case that is shown in Fig. \ref{fig:disp_FT} b), there are two mechanisms that destabilize the HSS: the homogeneous oscillatory mode with $q=0$ that originates from the calcium kinetics described by Eqs. (\ref{eq:chem}) and a wave-like perturbation at finite wavenumber ($q \ne 0$) induced by the mechanical feedback.
Above a critical mechanical coupling strength $F_T > F_T^{thr}>0$ the fastest growing mode (mode with largest real part of the eigenvalue) 
has a finite wavelength $q \ne 0$ and a nonzero imaginary part indicating wave dynamics.\\

In Fig. \ref{fig:disp_k} a) the influence of variation of the drag coefficient $\beta$ on the dispersion relation for fixed $F_T$ is shown. 
The wavelength $\Lambda_c=2\pi/q_c$ of the perturbation with largest wavelength increases with growing $\beta$ until the maximum with finite wavelength in the dispersion curve disappears. 
This discontinuity is visible in Fig. \ref{fig:disp_k} b) where $\Lambda_c$ is plotted versus the drag coefficient $\beta$ for different values of $F_T$. 
This figure also shows that the fastest growing wavelength $\Lambda_c$ decreases with the mechanical coupling strength. 
Nevertheless, the wavelength depends only weakly on $\beta$ over a large range of values. 

If a domain with finite size $L$ is considered, the dispersion relation becomes discrete since $q_n = n\pi/ L$ with $n$ integer has to be fulfilled. 
The first possible mode of a perturbation satisfying the boundary conditions has the wavelength $\Lambda_1=2L$. 
For small values of $\beta$, i.e. if the permeability is large, the maximum of the curve can be located at a smaller wavenumber than $q_1=\pi/L$. 
In this case the mode that dominates the growth of patterns is determined 
only by the system size and the global coupling approximation introduced in \citep{RAD10} applies.

\subsection*{Numerical simulations}

\begin{figure}[htbp]
   \begin{center}
      \includegraphics*[width=0.8\textwidth]{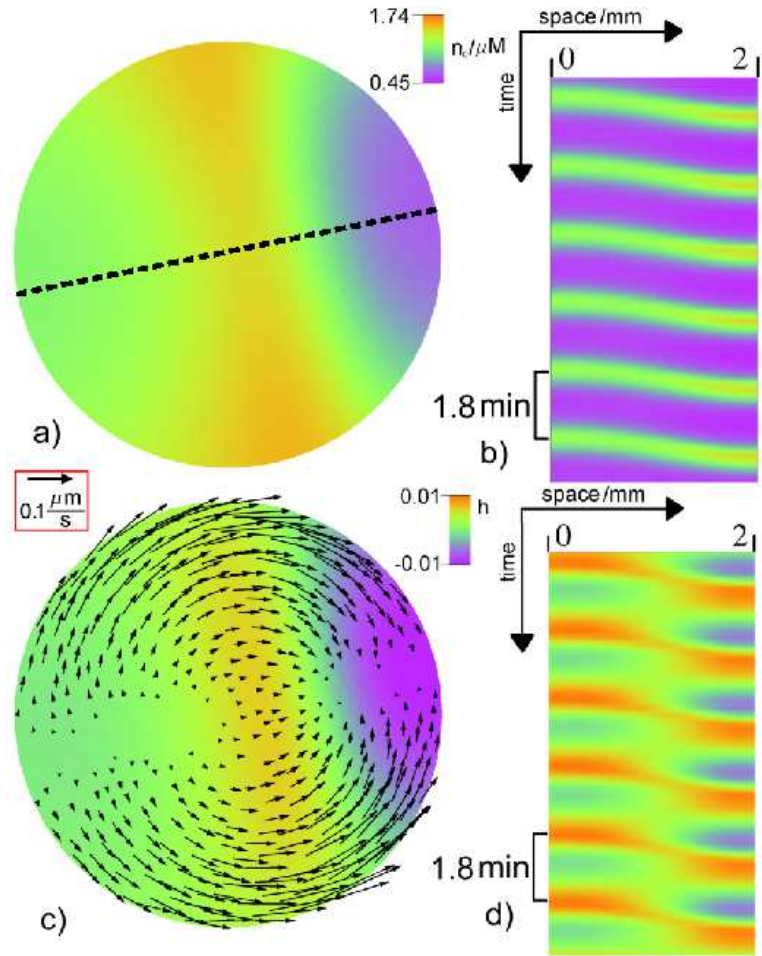}
      \caption{Traveling wave: a) Snapshot of the free calcium concentration $n_c$ and c) relative height field $h$ in color and the protoplasmic flow field $\rv{v}$ shown by arrows with length $\propto |\rv{v}|$. Space-time plot of $n_c$ b) and $h$ d) along the dotted line in subfigure a). The period of local 
      oscillations is $T=1.8$ $min$. The parameters are $F_T=18$ $kPa$ and $\beta=5.0\cdot 10^3$ $kg/(mm^3min)$. The remaining values can be found in Table \ref{tab:params}.}
      \label{fig:traveling}
   \end{center}
\end{figure}

\begin{figure}[htbp]
   \begin{center}
      \includegraphics*[width=0.8\textwidth]{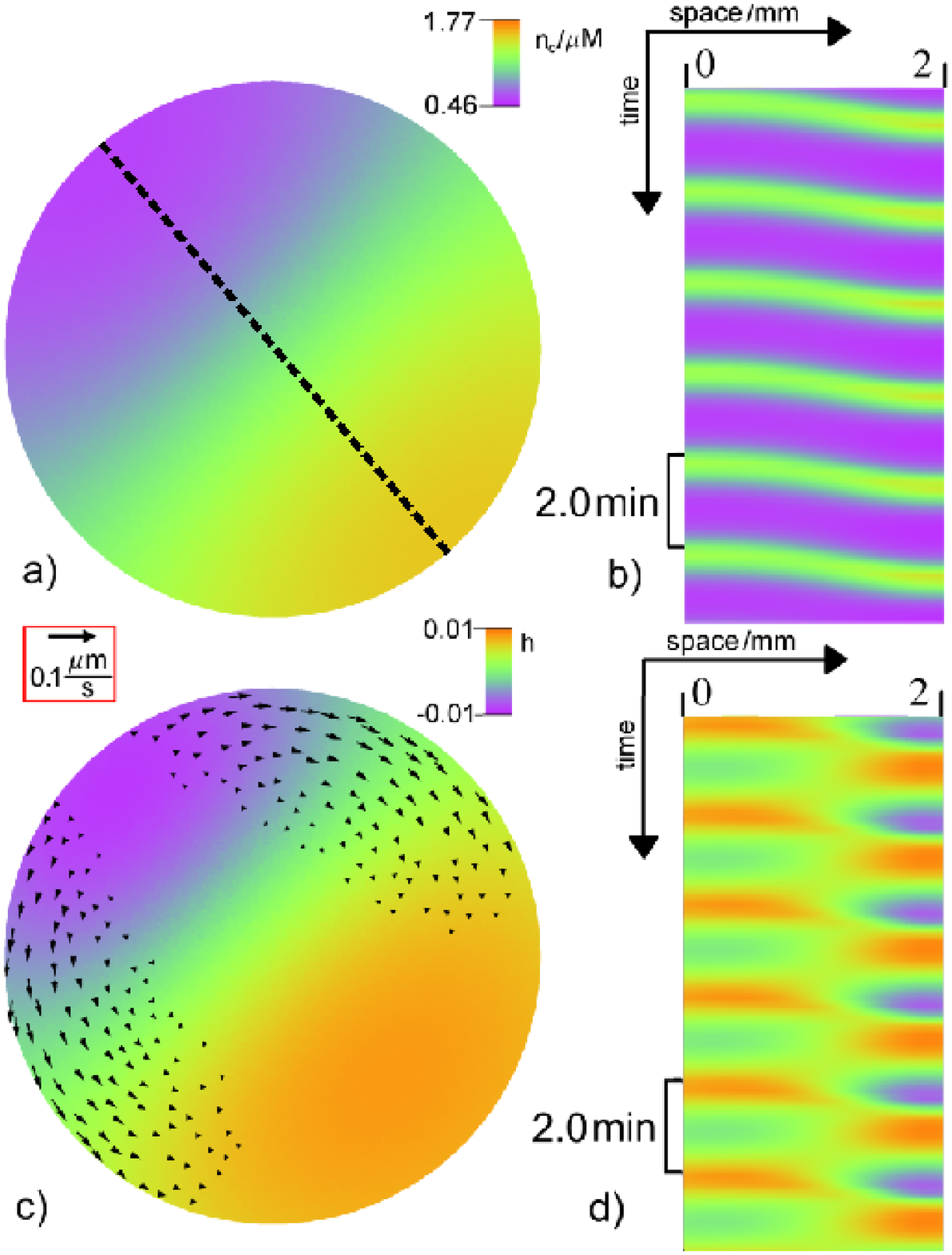}
      \caption{Standing wave: a) Snapshot of the free calcium concentration $n_c$ and c) relative height field $h$ in color and the protoplasmic flow field $\rv{v}$ shown by arrows with length $\propto |\rv{v}|$. Space-time plot of $n_c$ b) and $h$ d) along the dotted line in subfigure a). The period of local 
      oscillations is $T=2.0$ $min$. The parameters are $F_T=22$ $kPa$ and $\beta=5.0\cdot 10^4$ $kg/(mm^3min)$. The remaining values can be found in Table \ref{tab:params}.}
      \label{fig:standing}
   \end{center}
\end{figure}

\begin{figure}[htbp]
   \begin{center}
      \includegraphics*[width=0.8\textwidth]{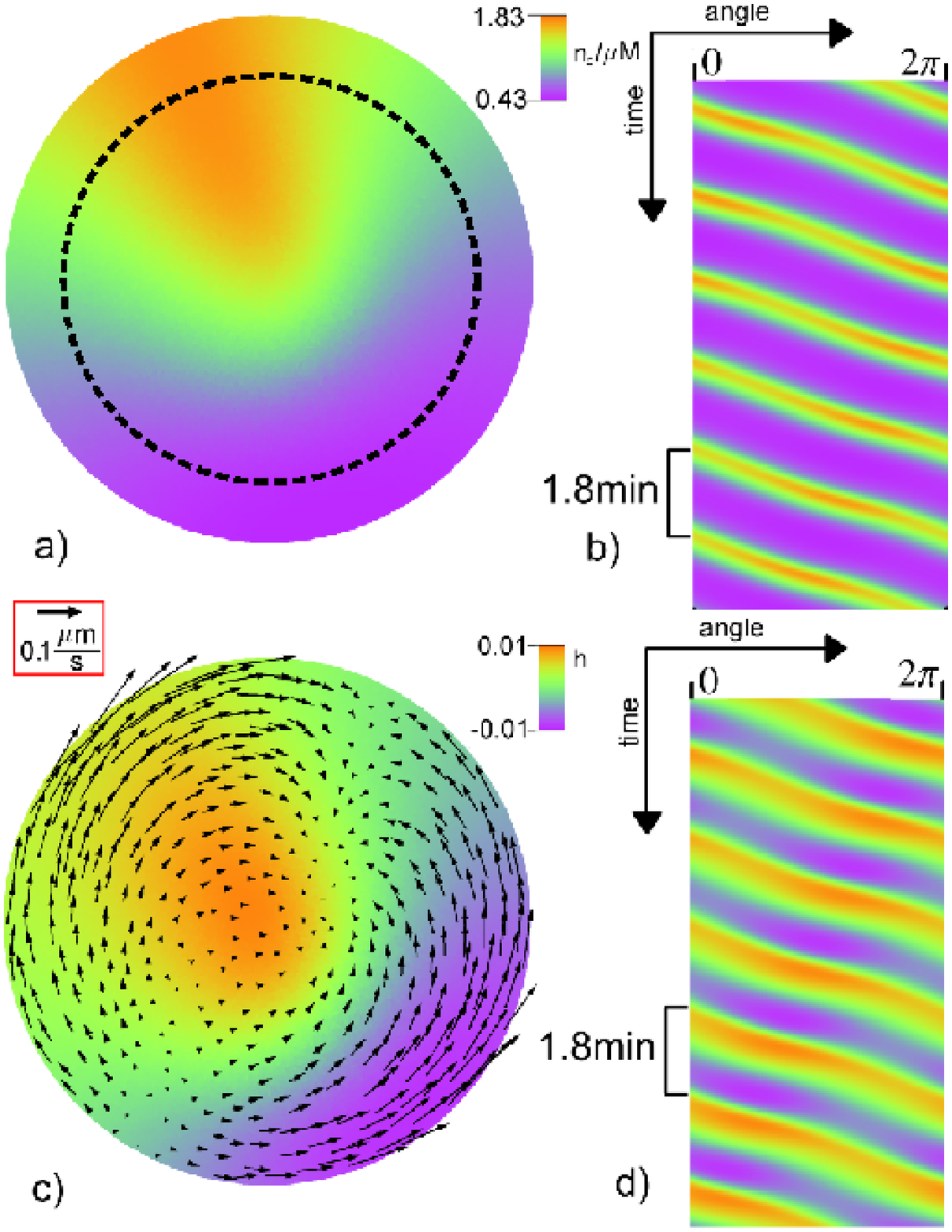}
      \caption{Single rotating spiral: a) Snapshot of the free calcium concentration $n_c$ and c) relative height field $h$ in color and the protoplasmic flow field $\rv{v}$ shown by arrows with length $\propto |\rv{v}|$. Space-time plot of $n_c$ b) and $h$ d) along a circle marked by the dotted line in a). The period of local oscillations is $T=1.8$ $min$.
      The parameters are $F_T=22$ $kPa$ and $\beta=5.0\cdot 10^4$ $kg/(mm^3min)$. For the remaining values, see Table \ref{tab:params}.}
      \label{fig:spiral}
   \end{center}
\end{figure}

\begin{figure}[htbp]
   \begin{center}
      \includegraphics*[width=0.8\textwidth]{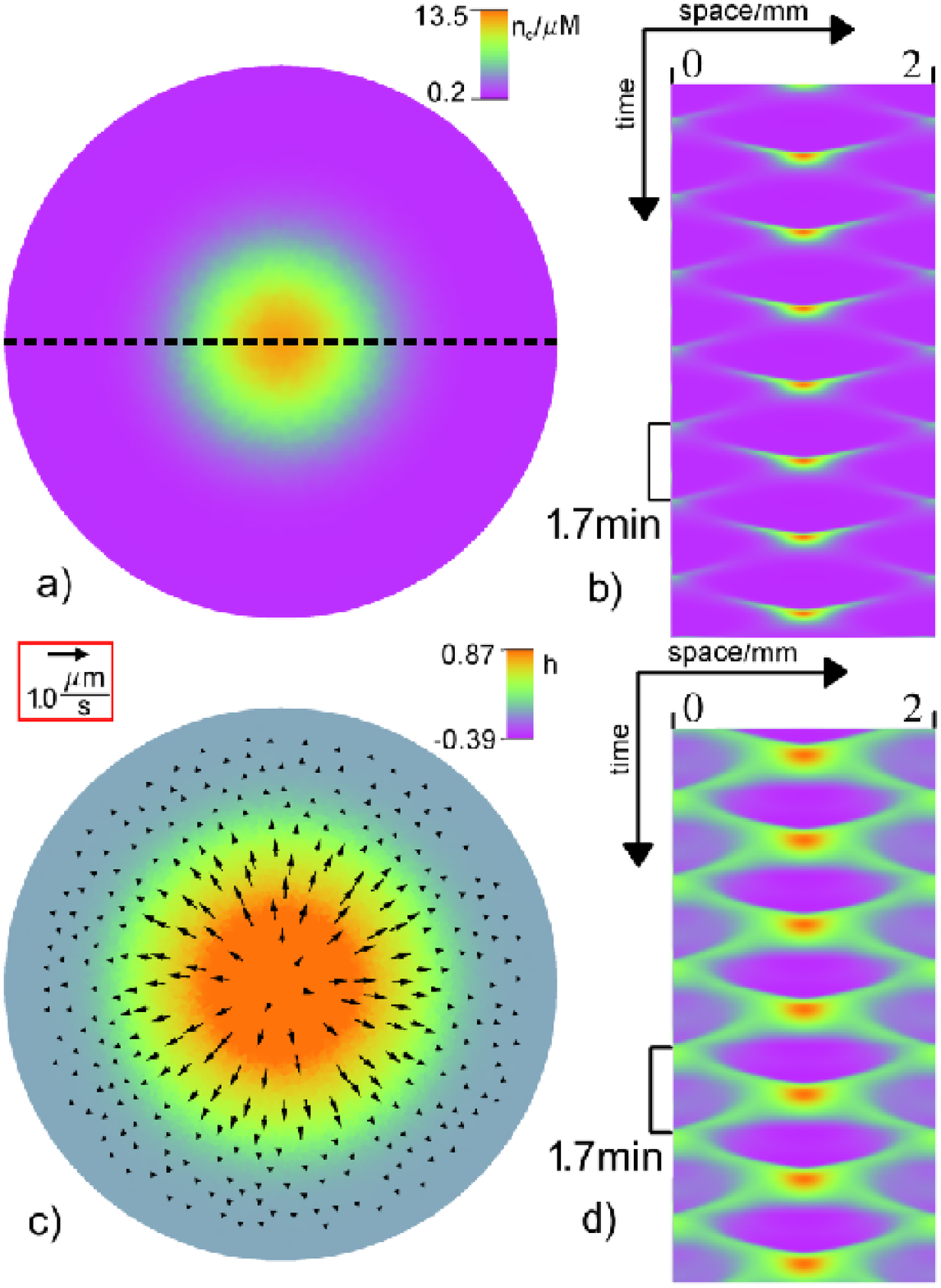}
      \caption{Radial wave: a) Snapshot of the free calcium concentration $n_c$ and c) relative height field $h$ in color and the protoplasmic flow field $\rv{v}$ shown by arrows with length  $\propto |\rv{v}|$. Space-time plot of $n_c$ b) and $h$ d) along the dotted line in a). The period of local 
      oscillations is $T=1.7$ $min$. The parameters are $F_T=194$ $kPa$ and $\beta=5.0\cdot 10^3$ $kg/(mm^3min)$. For the remaining values, see Table \ref{tab:params}.}
      \label{fig:circular}
   \end{center}
\end{figure}

\begin{figure}
   \begin{center}
      \includegraphics*[width=3.25in]{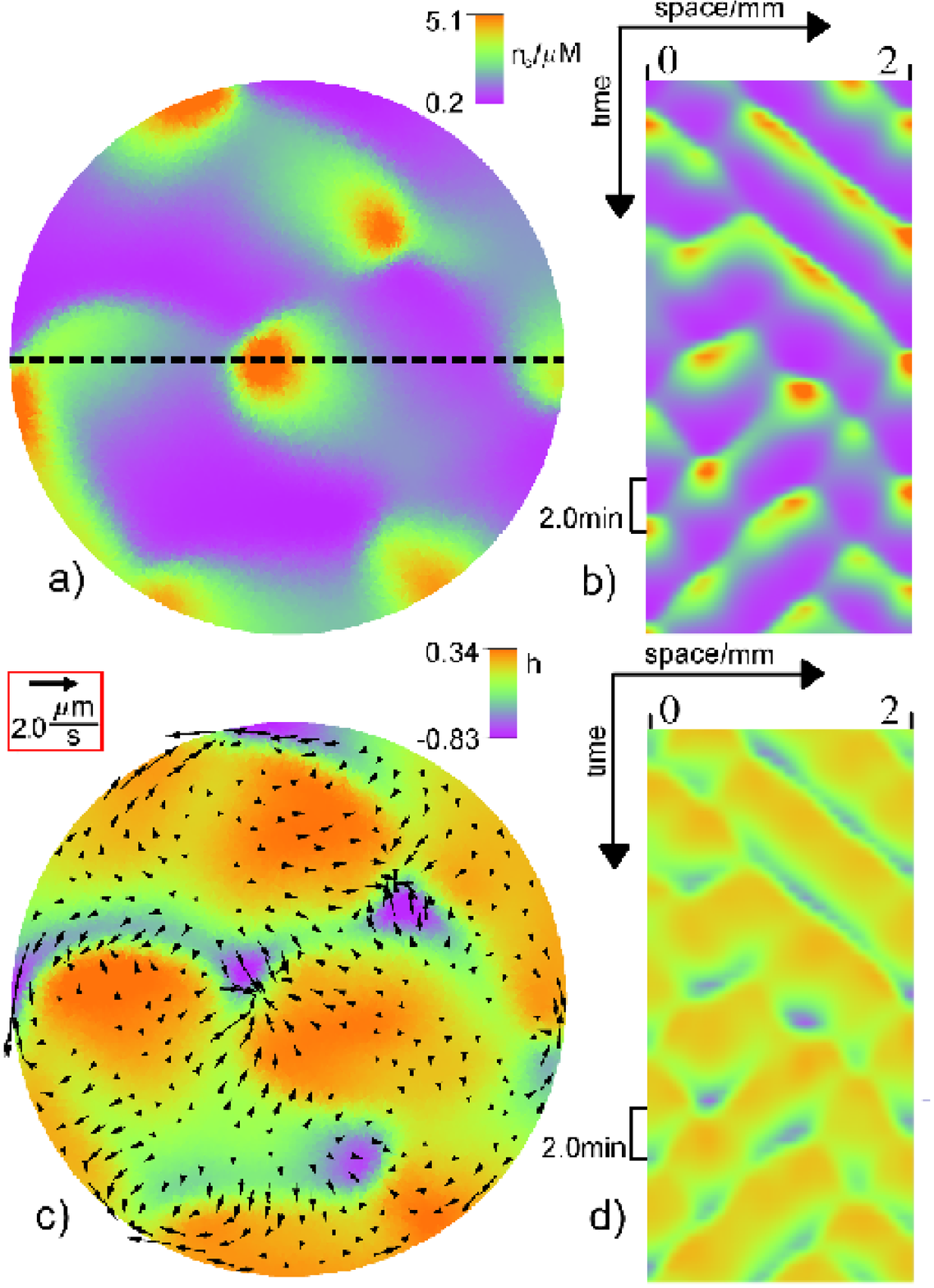}
      \caption{Irregular wave pattern: a) Snapshot of the free calcium concentration $n_c$ and c) relative height field $h$ in color and the protoplasmic flow field $\rv{v}$ shown by arrows with length $\propto |\rv{v}|$. Space-time plot of $n_c$ b) and $h$ d) along the dotted line in a). The parameters are $F_T=356$ $kPa$ and 
      $\beta=5.0\cdot 10^5$ $kg/(mm^3min)$. For remaining values, see Table \ref{tab:params}.}
      \label{fig:irreg}
   \end{center}
\end{figure}

For different values of the coupling strength $F_T$ and the drag coefficient $\beta$, we present numerical simulations of the dynamics of full nonlinear model equations. 
The corresponding phase diagram is shown in Fig. \ref{fig:phase_diag}. 
Therein, the black line separates the phase plane according to the shape of the dispersion curve obtained from the linear stability analysis of the HSS: It shows the threshold value $F_T^{thr}$ as a function of the
parameter $\beta$. 
When $F_T<F_T^{thr}$ there is no discrete wavenumber $q_k>0$, for which $\lambda(q_k)>\lambda(q_0=0)$. In the opposite case $F\geq F_T^{thr}$ there exists at least one $q_k>0$ with $\lambda(q_k)>\lambda(q_0)$. 
The prediction of homogeneous oscillations for $F_T<F_T^{thr}$ from linear stability analysis is confirmed by  numerical simulations for $\beta<10^5$ $kg/(mm^3min)$ (see Fig. \ref{fig:phase_diag}, blue discs). 
However, for larger values of $\beta$ one finds a considerable discrepancy between linear stability analysis and simulation results. 
For homogeneous oscillations in the calcium concentration the flow and deformation field both vanish.\\

Above $F_T^{thr}$, traveling, standing and spiral waves occur in the simulations. 
In Fig. \ref{fig:traveling} a traveling wave is depicted by snapshots and space-time plots. 
In addition to the free calcium concentration $n_c$ 
the plots show the relative height field $h$ and the sol velocity $\rv{v}$, since these quantities are most likely to be measured in experiments. 
We get an average velocity of the wave front that is 
about $0.086$ $mm/s$.\\

For values of $\beta$ around $5\cdot 10^4$ $kg/(mm^3min)$ standing waves are observed for the relative height field $h$, while for the concentration field $n_c$ a traveling wave appears (see Fig. \ref{fig:standing}).\\ 

Traveling and standing waves often coexist with single rotating spiral patterns. 
The pattern selection depends on the initial conditions, in particular on the number of phase singularities. 
Two counter-rotating spirals annihilate and give way to a traveling wave, whereas a single rotating spiral is stable. 
A single rotating spiral is presented in Fig. \ref{fig:spiral}. 
No distinct direction of propagation or rotation is preferred reflecting the symmetries of the system. 
The coexistence region of traveling waves and spirals is marked by violet squares in the phase diagram in Fig. \ref{fig:phase_diag}. 
From the space-time plots in Fig. \ref{fig:spiral} c) one finds a wave speed of $0.047$ $mm/s$.\\

For larger mechanical coupling strength $F_T$, there is no coexistence of spirals with traveling waves: a rotating wave is the only attractor (see Fig. \ref{fig:phase_diag}, green triangles). 
For large enough values of the drag coefficient $\beta$ a mode with largest possible wavenumber $\Lambda_1=2L$ determines the emerging patterns. 
This is confirmed by the numerical simulations. 
For $\beta<10^5$ $kg/(mm^3min)$ periodic patterns are obtained, even for large coupling $F_T$ (see Fig. \ref{fig:phase_diag}, brown squares). 
These patterns have a characteristic wavelength of the same order 
as the system size. 
The simplest pattern is an antiphase oscillations in the form of a radial wave that is reflected at the boundaries (see Fig. \ref{fig:circular}).\\ 

For large drag coefficients (see Fig. \ref{fig:irreg} and phase diagram \ref{fig:phase_diag}, pink circles), irregular patterns with a wavelength significantly smaller than the system size are obtained. 
For the wave segments in these spatiotemporal patterns we get a typical velocity of $0.03$ $mm/s$ that is much slower than that of traveling or spiral wave.\\

To address the question how much influence the advective coupling relative to diffusion has, one can consider a P\'eclet number that is given by the ratio of diffusive to advective time scales 
\begin{equation}
 Pe=\theta_{max}F_T/(D_c\beta),
 \label{eq:Pe}
\end{equation}
where $\theta_{max}\approx 0.01$ is the amplitude for the oscillations in the variable $\theta$ in Eq. (\ref{eq:theta}). 
%
%
In Fig. \ref{fig:phase_diag} the blue line corresponds to $Pe=1$. 
Note, that above the line $Pe > 1$, there are no homogeneous oscillations and different types of patterns prevail. 
In some cases, simple patterns like rotating spirals are traveling waves are also found for $Pe < 1$. 
Altogether, this consideration shows that the mechanical coupling has to be strong enough to overcome the homogenizing effect of diffusion for mechanochemical waves and patterns to emerge. 


\section*{Discussion}
In this article we have combined a novel mechanical continuum model of the cytoplasm as a poroelastic active medium with the Smith-Saldana model for calcium oscillations in Physarum protoplasma \citep{SMI92}. 
Upon increase of the mechanochemical coupling strength the homogeneous steady state in this model are destabilized by an oscillatory Turing instability with finite wave number. 
Homogeneous oscillations of the calcium concentration are replaced by spatiotemporal patterns and waves connected with local deformation and 
fluid motion. 
Practically all experimentally observed spatiotemporal deformation patterns and \citep{TAK08,TAK10} including rotating spirals, traveling and standing waves, antiphase oscillation and irregular, chaotic waves could be reproduced by numerical simulations of this model. 

In contrast to earlier models (see e.g. \citep{TSU11}) for Physarum protoplasma a closed set of mechanical force-balance equations is derived that allows for explicit computations of pressure and flow fields. 
The cytoplasm is treated as a two-phase material, consisting of a passive fluid sol phase and an active solid viscoelastic gel phase. 
The basic idea was already sketched in \citep{RAD10}, where however only the limiting case where the mechanochemical coupling can be approximated by a global coupling in the reaction-diffusion dynamics for the calcium dynamics. 
Here, we have instead analysed and simulated the interplay of mechanical deformation of the cytoskeleton, fluid flow in the cytosol and chemical (= calcium) concentration and the associated spatiotemporal dynamics for a wide range of parameters.\\

Unlike other models that treat the cytoskeleton as a viscoelastic fluid (see e.g. \citep{CAJ11}) we consider the cytoskeleton to be a viscoelastic solid described the Kelvin-Voigt model.
This is a valid approximation in Physarum, since the relaxation time of elastic tension is larger by about a factor of three than the calcium oscillation period in the system \citep{NAG78}. 
The porosity of the solid gel phase allows a flow of cytosol that carries calcium and regulates the tension generation. 
This is represented by an advectice transport of calcium in addition to diffusion and introduces a nonlinear feedback mechanism that couples flows and deformations to a nonlinear reaction kinetics. 
Related models of active gels and fluids \citep{JOA07,BAN10,BOI11} consider the transport of motors, whereas we have considered the transport of calcium that acts as a motor-regulating species in Physarum. 
We have limited our considerations to small displacements $\rv{u}$ and deformations $|\vn\rv{u}|$.
Thus, the equations are formulated up to linear order in the displacement field and its gradient. 
\\
A linear stability analysis of a homogeneous steady state with constant calcium concentration and zero deformation and flow has been carried out for the presented model. 
The resulting dispersion relations reveal that the mechanical feedback provides a new mechanism of an oscillatory Turing instability with nonzero wavenumber if the mechanical coupling strength is positive $F_T>0$. 
In our numerical simulations, homogeneous oscillations get destabilized for sufficiently large coupling strength $F_T$. 

Rotating spirals and traveling waves obtained in the simulations are common patterns found in experiments
\citep{TAK08,TAK10,SH10}. 
The wave speed we obtain for traveling and spiral waves agree with the findings in \citep{TAK08}. 
The irregular pattern in Fig. \ref{fig:irreg} resembles the experimentally observed one in Fig. \ref{fig:exp} b). 
The pattern shown in Fig. \ref{fig:circular} has the same symmetry as the antiphase oscillation in experiments (Fig. 1d, \citep{TAK08}).  
Note, however, that both, the irregular and the antiphase pattern, were obtained at large coupling strength $F_T$ yielding also large deformations. 
Thus, these results violate the small deformation assumption of our model and should be used only for qualitative comparisons. 
An even better agreement between simulations and experiments is obtained if a model with a softer periphery of the droplet is assumed \citep{rad-phd-2013}. 
There, antiphase oscillations are already found for smaller values of $F_T$ within the range of the validity of the linear elasticity approach used. 
The mechanism of the experimentally observed transition between different patterns on a timescale much larger than the typical calcium oscillation period is not fully understood. 
The phase diagram in Fig. \ref{fig:phase_diag} shows that a 
minor change in the coupling strength $F_T$ or drag coefficient $\beta$ can result in a different type of pattern. 
Hence, a variation of one of these parameters over the time of the experiment may be responsible for qualitative changes.\\

In the presented model we consider fixed boundaries (Dirichlet conditions), which is a good approximation when the model considers the first hours after dissection of the droplet. 
Modeling the stage of migration of Physarum droplet would requires a modification of the model allowing consideration of dynamics on long time scale e. g. fluidization of the cytoskeleton.
An option is to use the Maxwell model of viscoelasticity. 
Moreover, free-boundary conditions have to be imposed and an interaction with the substrate must be considered (see e.g. \citep{ALT99}). 
Such an extension of the model may eventually help to understand the interplay of chemical and mechanical processes in the self-organized amoeboid movement of Physarum droplets. 

\begin{table}[H]
 \begin{center}
    \begin{tabular}{|l|l|l|}
     \hline
      Parameter & Value & Description\\
     \hline
     \hline
     \multicolumn{3}{|c|}{Chemical parameters} \\
     \hline
      $k_L$ & $0.24 min^{-1}$ & leaking rate of vacuoles \citep{SMI92}\\
     \hline
      $k_V$ & $4.8 min^{-1}$ & pumping rate of vacuoles \citep{SMI92}\\
     \hline
      $k_Q^0$ & $60 min^{-1}$ & max. phosphorylation rate of MLCK \citep{SMI92}\\
     \hline
      $k_E$ & $6.0 min^{-1}$ & dephosphorylation rate of MLCK \citep{SMI92}\\
     \hline
      $k_P$ & $30 min^{-1}$ & phosphorylation rate of $LC_1$ \citep{SMI92}\\
     \hline
      $k_D$ & $12 min^{-1}$ & dephosphorylation rate of $LC_1$ \citep{SMI92}\\
     \hline
      $K^\star$ & $1.5 \mu M^{-1}$ & effective activation constant for the\\
      & & AC-cAMP-PKA chain \citep{SMI92}\\
     \hline
      $K_a$ & $2.3 \mu M^{-1}$ & \Ca affinity with dephosphorylated MLCK \\
     \hline
      $K_b$ & $0.15 \mu M^{-1}$ & \Ca affinity with phosphorylated MLCK \citep{SMI92}\\
     \hline
      $N_c$ & $25 \mu M$ & equilibrium total calcium concentration \citep{SMI92}\\
     \hline
      $N_M$ & $10 \mu M$ & total myosin concentration \citep{SMI92}\\
     \hline 
      $\tau_T$ & $0.2 min$ & relaxation time for tension generation \citep{RAD10} \\
     \hline
     \hline
     \multicolumn{3}{|c|}{Mechanical parameters} \\
     \hline
      $F_T$ & $0-350 kPa$ & mechanical coupling strength \citep{NOR40}\\
     \hline        
      $D_c$ & $0.03 \frac{mm^2}{min}$ & free calcium diffusion coefficient \citep{DON87} \\
     \hline
      $K$ & $8.9 kPa$ & gel compression modulus \citep{NOR40,BOR04} \\
     \hline
       $G$ & $8.9 kPa$ & gel shear modulus \citep{NOR40,BOR04} \\
      \hline
       $\eta_{\textrm{sol}}^{shear}$ & $1.0 Pas$ & effective sol shear viscosity \citep{SAT83,BYK09} \\
      \hline
       $\eta_{\textrm{gel}}^{shear}$ & $1.0 Pas$ & effective gel shear viscosity; indirectly from \citep{PEL09}\\
      \hline
      $\eta_{\textrm{sol}}^{bulk}$ & $0.0Pas$ & effective sol bulk viscosity \\
      \hline
      $\eta_{\textrm{gel}}^{bulk}$ & $0.0Pas$ & effective gel bulk viscosity \\
      \hline
       $\beta$ & $5\cdot 10^2-5\cdot 10^5$ & drag coefficient, related to pore size \citep{NAG75,BRX87} \\
       & $kg/(mm^3min)$ &  \\
      \hline     
       $\sol^0$ & $0.75$ & sol volume fraction \citep{KES76} \\
     \hline
    \end{tabular}
 \end{center}
 \caption{Standard parameter set with description and references}
 \label{tab:params}
\end{table}


\subsection*{Acknowledgements}
We acknowledge financial support from the German Science Foundation (DFG) within the GRK 1558 ``Nonequilibrium Collective Dynamics in Condensed Matter and Biological Systems''. 
Furthermore, we thank J.R. Shewchuk for providing the free Software \textit{Triangle} we used for generation of two-dimensional FEM meshes. We are indebted to S. Alonso, C. Bernitt, M. J. B. Hauser, 
T. Nakagaki, U. Strachauer, and T. Ueda for stimulating and useful discussions. 


\bibliographystyle{biophysj}


%
%
%
%
%
%
%
%
%
%
%
%
%
%
%
%
%
%
%
%

%

\end{document}


\maketitle

\clearpage

%


\section*{Derivation of the Mechanical Equations}

\subsection{A continuity equation for the sol}
In the model for Physarum protoplasmic droplets we assume a sponge-like material filled with a fluid that is transported through the pores (see Fig. 2 in the main article). 
In regions where the influx of sol exceeds the outflux, the sponge will expand. 
As a result the sol fraction increases. 
It is a basic assumption in our model that the solid (gel) fraction in the body-reference coordinate frame is constant in space and time: $\gel=\gel^0=const.$ 
Given the deformation gradient $\rv{F}=\vn_{\rv{x}}\rv{X}=\mathbf{1}+\vn\rv{u}$, the sol/gel fractions $\varrho_{sol/gel}$  in the lab frame are 
\begin{equation}
 \begin{array}{lll}
  \sol & = & \varrho_{sol}det\rv{F} \\
  \gel & = & \varrho_{gel}det\rv{F}.
 \end{array}
\end{equation}
The relation $\varrho_{sol}+\varrho_{gel}=1$ holds exactly in the lab frame, while in the reference frame we have:
\begin{equation}
 \sol+\gel=det\rv{F}=1+\vn\cdot\rv{u}+\mathcal{O}(\vn\rv{u}^2).
\end{equation}
Since $\gel$ is a constant we arrive at the simple relation (with $\sol^0:=1-\gel^0$)
\begin{equation}
 \sol = \sol^0+\vn\cdot\rv{u}+\mathcal{O}(\vn\rv{u}^2).
 \label{eq:rho_sol}
\end{equation}
To linear order in $\vn\rv{u}$ (small strains) the incompressibility condition reads (see also \citep{RAD13} and \citep{rad-phd-2013})
\begin{equation}
 \vn\cdot (\sol^0\rv{v}+\gel^0\dot{\rv{u}})=0.
 \label{eq:inc}
\end{equation}
Using Eqs. (\ref{eq:rho_sol}) and (\ref{eq:inc}) a continuity equation of the form 
\begin{equation}
 \dot{\rho}_{sol}+\vn\cdot((\rv{v}-\dot{\rv{u}})\sol) = 0
\end{equation}
is derived that is valid in linear order in $\vn\rv{u}$. 
With this we show the analogy to other models of two-phase flow \citep{ALT99,COG10}. 
Here, we neglect sol-gel transformations. 
For the presented mechanical model we only need the zeroth order in $\rho$ since higher order corrections lead to second order terms of $\vn\rv{u}$ in the 
force balance equations (see supplementary material in \citep{RAD13}).

\subsection*{Derivation of the Force-Balance Equations}
The sol and gel stresses are divided into a dissipative and non-dissipative part:  $\rv{\sigma}_{gel}=\rv{\sigma}_{gel}^{non}+\rv{\sigma}_{gel}^{dis}$ and $\rv{\sigma}_{sol}=\rv{\sigma}_{sol}^{non}+\rv{\sigma}_{sol}^{dis}$. 
First we consider only the dissipative part, write a functional for the entropy production rate and minimize that functional compare also e.g. \citep{ALT99}.
This assumes local thermodynamic equilibrium. 
We use the summation convention and write the functional: 
\begin{equation}
\begin{array}{ll}
  -T\dot{S} & =J^{dis}[\dot{\rv{u}},\rv{v},p]= -\frac{1}{2}\int_{\mathfrak{B}}d\rv{x}(\gel^0\sigma_{gel \alpha\beta}^{dis}\partial_\beta\dot{u}_\alpha+\sol^0\sigma_{sol \alpha\beta}^{dis}\partial_\beta v_\alpha \\
   & +\beta\gel^0\sol^0(\dot{u}_\alpha-v_\alpha)(\dot{u}_\alpha-v_\alpha)-2p(\gel^0\partial_\alpha\dot{u}_\alpha+\sol^0\partial_\alpha v_\alpha).
\end{array}
\end{equation}
The minimization is carried out under the constraint of incompressibility, leading to an additional hydrostatic pressure field $p$. 
The quantities $\rho_{sol/gel}$ are treated as constants (see above). 
We consider the situation where $|\vn\rv{u}|\ll 1$.
The resulting Euler-Lagrange equations are:
\begin{align}
  \frac{\partial j^{dis}}{\partial\dot{u}_\gamma} -\D{\nu}(\frac{\partial j^{dis}}{\partial(\D{\nu}\dot{u}_\gamma)})& =0 \label{eq:lagrange_1}\\
  \frac{\partial j^{dis}}{\partial v_\gamma} -\D{\nu}(\frac{\partial j^{dis}}{\partial(\D{\nu}v_\gamma)})& =0 \label{eq:lagrange_2}\\
  \frac{\partial j^{dis}}{\partial p} -\D{\nu}(\frac{\partial j^{dis}}{\partial(\D{\nu}p)})& =0\label{eq:lagrange_3}.
\end{align}
Linear constitutive laws for an isotropic viscous medium are used that involve the shear and bulk viscosities of sol and gel phase:
\begin{equation}
 \begin{array}{ll}
  \sigma_{gel \alpha\beta}^{dis}= & \eta_{gel}^{shear}(\D{\beta}\dot{u}_{\alpha}+\D{\alpha}\dot{u}_{\beta}-\frac{2}{d}\D{\lambda}\dot{u}_{\lambda}\delta_{\alpha\beta})+\eta_{gel}^{bulk}\D{\lambda}\dot{u}_{\lambda}\delta_{\alpha\beta} \\
  \sigma_{sol \alpha\beta}^{dis}= & \eta_{sol}^{shear}(\D{\beta}v_{\alpha}+\D{\alpha}v_{\beta}-\frac{2}{d}\D{\lambda}v_{\lambda}\delta_{\alpha\beta})+\eta_{sol}^{bulk}\D{\lambda}v_{\lambda}\delta_{\alpha\beta} \\
  \label{eq:vstress}
 \end{array}
\end{equation}
We keep the spatial dimension $d$ as a variable in the supplementary material. 
To obtain the equations in the main text one has to set $d=2$. Eqs. (\ref{eq:lagrange_1}) and (\ref{eq:lagrange_2}) yield the force balance equations
\begin{align}
  -\beta\sol\gel(\dot{u}_\gamma-v_\gamma)+\gel\D{\nu}(\sigma_{gel \gamma\nu}^{dis}-p\delta_{\gamma\nu}) & =0 \label{eq:fb_1}\\
  \beta\sol\gel(\dot{u}_\gamma-v_\gamma)+\sol\D{\nu}(\sigma_{sol \gamma\nu}^{dis}-p\delta_{\gamma\nu}) & =0 \label{eq:fb_2},
\end{align}
while Eq. (\ref{eq:lagrange_3}) reproduces the incompressibility condition.\\
So far, we only included viscous stresses.
However, the gel phase should be active and elastic. 
The Kelvin-Voigt model of viscoelasticity is incorporated by replacing the purely viscous gel stress discussed so far by the sum of viscous elastic and active stress: $\rv{\sigma}_{gel}^{dis}\rightarrow \rv{\sigma}_{gel}^{dis}+\rv{\sigma}_{gel}^{el}+\rv{\sigma}_{gel}^{act}$. \\
The general linear isotropic constitutive law for an elastic solid is:
\begin{equation}
 \sigma_{gel \alpha\beta}^{el}= G(\D{\beta}u_{\alpha}+\D{\alpha}u_{\beta}-\frac{2}{d}\D{\lambda}u_{\lambda}\delta_{\alpha\beta})+K\D{\lambda}u_{\lambda}\delta_{\alpha\beta} \\
 \label{eq:estress}
\end{equation}
with shear modulus $G$ and compression modulus $K$.
The active stress generated by the actomyosin system is also considered as isotropic:
\begin{equation}
 \sigma_{gel \alpha\beta}^{act}=T_a\delta_{\alpha\beta}.
 \label{eq:astress}
\end{equation}
Summing Eqs. (\ref{eq:vstress}),(\ref{eq:estress}) and (\ref{eq:astress}) up to the total gel stress and inserting it into Eq. (\ref{eq:fb_1}) one obtains Eq. (9) in the main article; Eq. (\ref{eq:fb_2}) corresponds to Eq. (10) in the main article.

\subsection*{Osmotic swelling pressure}
It was stipulated in \citep{OST84} that in a highly concentrated polymer solution like cytoplasm there should by an additional gel pressure $p_{osm}=f(\varrho_{gel})$ that causes a swelling with increasing concentration $p\rightarrow p+p_{osm}$.
In our model the osmotic stresses due to differences in the concentration of the actin phase are neglected. The following considerations will justify this step. 
In the framework of the small strain theory we can expand: 
\begin{equation}
 p_{osm}(\varrho_{gel})\approx p_{osm}(\gel^0)-\frac{\partial f}{\partial\varrho_{gel}}(\gel^0)\gel^0\partial_\alpha u_\alpha.
\end{equation}
We add this to the hydrostatic pressure in Eq. (\ref{eq:fb_1}) and get the following terms:
\begin{equation}
 ...+G\partial_{\alpha\alpha}u_\beta+(K+\frac{d-2}{2}G)\partial_{\alpha\beta}u_\alpha+\frac{\partial f}{\partial\varrho_{gel}}(\gel^0)\gel^0\partial_{\alpha\beta}u_\alpha+\partial_\beta(T_a-p)+...
\end{equation}
It becomes obvious that one can absorb the osmotic pressure term in the compression modulus:
\begin{equation}
 K'=K+\frac{\partial f}{\partial\varrho_{gel}}(\gel^0)\gel^0.
\end{equation}
Consequently, including this form of osmotic pressure does not lead (in the linear approximation) to any behavior that is not captured by the model presented here.


\bibliographystyle{biophysj}